\documentstyle[12pt]{article}

\newcommand{\ue}%
{\mbox{c\hspace{-0.4em}\rule[ 0.5ex]{0.3em}{0.04ex}\hspace{0.1em}}}
\newcommand{\Ue}%
{\mbox{C\hspace{-0.6em}\rule[ 0.75ex]{0.4em}{0.04ex}\hspace{0.2em}}}
\newlength{\signlength}%
\newcommand{\gs}%
{\settowidth{\signlength}{$<$}%
\raisebox{-0.35\signlength}%
{\makebox[1.7\signlength][c]{\makebox[0.pt]{$\sim$}%
\raisebox{0.6\signlength}{\makebox[0.pt]{$>$}}}}}
\newcommand{\ls}%
{\settowidth{\signlength}{$<$}%
\raisebox{-0.35\signlength}%
{\makebox[1.7\signlength][c]{\makebox[0.pt]{$\sim$}%
\raisebox{0.6\signlength}{\makebox[0.pt]{$<$}}}}}
\renewcommand{\ge}%
{\settowidth{\signlength}{$>$}%
\setlength{\unitlength}{0.1\signlength}%
\mbox{\begin{picture}(17,7.5)%
\linethickness{0.045\signlength}%
\put(0.0,1){\makebox(17,7.5){$>$}}%
\multiput(4.5,-1)(0.25,0.125){32}{\line(1,0){0.25}}%
\end{picture}}}
\renewcommand{\le}%
{\settowidth{\signlength}{$>$}%
\setlength{\unitlength}{0.1\signlength}%
\mbox{\begin{picture}(17,7.5)%
\linethickness{0.045\signlength}%
\put(0.0,1){\makebox(17,7.5){$<$}}%
\multiput(12.,-1)(-0.25,0.125){32}{\line(-1,0){0.25}}%
\end{picture}}}
\title{Three component model of pseudogap state for high temperature
superconductors}
\author{Sergeeva G.G $^{1}$, Vakula V.L.$^{2}$ }
\date{}
\begin{document}
\maketitle
\thispagestyle{plain}

\begin{center}

$^{1}$ National Science Center "Kharkov Institute of Physics

and Technology", Academicheskaya st. 1, 61108, Kharkov, Ukraine

$^{2}$  B.Verkin Institute for Low Temperature Physics and Engineering of the

National Academy of Sciences of Ukraine, 47 Lenin Ave., Kharkov
61103, Ukraine

\end{center}

\begin{abstract}
	For underdoped cuprates HTS three component model of pseudogap
state is proposed where it is shown that at $T<T^{*}$
in the addition to Jahn-Teller
small polarons (with total spin 1/2) and holes the zero
dimensional superconducting fluctuations  are developing in copper-oxygen
planes. These fluctuations are the local "hole - Jahn Teller small polaron"
pairs with BCS pairing. At $T<T^{*}$ ( $T^{*}$ is the pseudogap
temperature) the crossover of three-dimensional charge to "two-dimensional"
one occurs, and the polaron shift of the energy leads to the compensation of
Coulomb repulsion between polaron and hole and to their on-site attraction in
copper-oxygen planes. Some experimental evidences of the local
"hole - polaron" pairing at $T<T^{*}$  are discussed.

\end{abstract}

1. Despite of the intensive researches of the nature of high
temperature superconductivity (HT superconductivity) the questions
about pairing mechanism and about the nature and number of carriers
remain open. Today for normal state of underdoped high temperature
superconductors (UD HTS)  the two-component model of carriers we can
consider as firmly established fact: at $T>T^{\ast}$ the small
polarons and holes are the heavy and light carriers respectively. Here
$T^{\ast}$ is the temperature of the transition in pseudogap state (PGS) when
$Cu O_2$ planes are in stripe state. For normal state  ones of the first
evidences of coexisting these two carrier types were the
measurements of optical conductivity for UD HTS [1-3]. Later the
spin susceptibility measurements for $La _{2-x} SR_ x Cu O_4$ [4] let to
determine the doping dependence of the part for each carrier, but the polaron
state ( it is polaron or bipolaron) was undetermined. This question for
cuprates HTS is of fundamental importance because namely the
existing of Jahn Teller small polarons (JT SP) for doped antiferromagnets
(AF) with large values of dielectric constant and strong oscillations of
oxygen ions [5] was initial point for HTS searches. At the doping the extra
holes are localized on the transition metal ions that leads to the change
of their valence and to strong JT distortions.  At the doping increase and
at the temperature decrease the transition of AF into metal occurs with two
carrier types and with the of principle possibility of superconducting
transition (SCT) with enough high value of $T_c$.

Now it is clear that the understanding of the PGS nature at
$T^{\ast} >T>T_c$ furnish the clue to HT superconductivity. Next two
circumstances are the reason for this supposition:  1) the change of
the density of state  begin at $T \sim T^{\ast}$ and goes on up to $T_c$;
2) at $T= T_c$  coherent superconducting state is forming
 but noticeable change of the density of state does not
observe (see refs. in [6]). Under BEC theory for polarons in
 ref.[6] three component model of PGS was proposed in which bipolarons
were the third component.  In this paper it is discussed the supposition
that  at $T\le T^{\ast}$  the BCS pairing states of hole and JT SP are the third
component for PGS.

2. For UD HTS the coexisting of JT SP and holes at  $T > T^{\ast}$
stimulates the interest to the studying the possibility of their
pairing. In Refs.[7-8] the possibility of such pairing was shown.
At that for HTS the mechanism of the suppression of on-site Coulomb
repulsion  $U$ for two particles is the main problem.
At the first E.K.Kudinov [8] for two component model
of carriers show the principle possibility of their on-site attraction: i)
JT SPs lead to the band narrowing and to the polaron shift of the energy
$E_p = g_{JT}^{2}/2M\omega^{2}$, ii) at $(-E_p +U)<0 $ both as the
compensation of Coulomb repulsion and so
the on-site attraction between the hole and JT SP take place. At that
all many particles interactions between the hole and JT SP are exponential
renormalizated. Here $g_{JT}$ is a elastic constant of JT interactions
(JTI) of holes
with oscillating oxygen ions. JT SP is a hole in $CuO_2$ plane bounded by
means JTI with complex of two neighboring $Cu_{+2}+4 O_{-2}$ "square" with
common oxygen ion. The the diagonals of these squares are distorted by
$Q_2$ phonon mode. Total spin of JT SP is equal 1/2, and spins of two
$Cu^{+2}$ are antiparallel (a polaron with parallel spins of two $Cu^{+2}$
is bound three spin polaron bound [9]). As it shown by Kudinov
the pairing of hole and JT SP in model BCS leads to SCT in $CuO_2$
plane with the temperature $T_{cr} \sim |E_p - U|$.  At that the pair
"hole-JT SP" is localized by the $Cu-O$ complex of JT SP and is local pair
with correlation length $\xi_{ab} \le 2R_{Cu-O}$ ($R_{Cu-O}$ is the distance
between $Cu^{+2}$ and $O^{-2}$ in $CuO_2$ plane).

Exponential renormalization of all interactions
leads to exponential small contribution in energy of bipolaron pairing
relatively with the contribution of the  "hole-JT SP" pairs[8]. This means
that the bipolaron pairing is realized but for SCT it is not important. But
the bipolaron pairing of bound three spin polarons can has crucial meaning
in first stage of the stripe forming [10].

3.For UD HTS at incoherent interlayer tunnelling the charge transfer
along $c$-axis is the result of thermal fluctuations
$k_B T >t^{2}_c/ t_{ab}$, where $t_c$ and  $t_{ab}$ are the probability
of the charge transfer along $c$-axis and in $Cu O_2$ plane, $k_B$ is
the Boltzman constant. At the lowering temperature the thermal
fluctuations cut down the interlayer tunnelling, and at
$ T^{*}=t_c ^{2}(T^{*})/ k_B t_{ab}$ the crossover from three-dimensional
charge to two-dimensional one occurs. At $T^{\ast} \sim T_{cr}$ this
dimensional crossover is accompanied by pairing JT SP and holes
in $Cu O_2$ plane. The measurements of optical conductivity are
the convincing evidence of such pairing where it is shown that at
$ T^{*} $ the  $c$-axis component of electronic kinetic energy
and carriers mass twice increase [11]. The lowering
temperature $T<T^{\ast}$ leads to the growth of number "hole-JT SP" pairs
$n_{hp}$ ( the distance between which is greater than their correlation
length $\xi_{ab}$), i.e. to zero-dimensional (0D) superconducting
fluctuations (SCF) [12-13].  Further lowering temperature leads to
the increasing of $n_{hp}$ and $\xi_{ab}$, so that at big enough
$\xi_{ab}$ local pairs begin overlap and
the two transitions occur: at first to two-dimensional
(2D) SCF, and at more low temperature to three-dimensional
(3D) SCF [12-14].  At that holes number, $n_h$, and polarons number,
$n_p \ge n_h$, are decreasing that is according with the observation
of the absence of noticeably change
density of states at $T_c$ [6]. For example, at the Hall
effect measurements for $YBa_2 Cu_3 O_{6+x} (T_c = 87.4K)$ was found out that
$n_h$ decreases twice at lowering the temperature from $240K$ up to $100K$:
$n_h(240K)\sim 5.4 \cdot10^{21}cm^{-3}$, and $n_h (100K)\sim 2.7 \cdot10^{21}
cm^{-3}$ [15].

The transition to 2D SCF with the dependence
$\xi_{ab}(T)= \overline{\xi}_{ab}(T/T_{2D} - 1)^{-1/2}$ leads to the
semiconducting dependence of the $c$-axis resistivity
with the probability of the charge transfer which is depending on
the temperature [14]
\begin{equation}\label{1}
t_c(T)\approx\overline{\xi}_c ^{2}/ \overline{\xi}_{ab} ^{2}
(T/T_{2D} - 1),
\end{equation}
where $T_{2D}$ is the temperature of the two-dimensional SCT for the
isolated $Cu O_2$ plane,
$\overline{\xi_c}$, $ \overline{\xi}_{ab} $ are the values of the
correlation lengths at $ T_{2D}$, and $E_F$ is Fermi energy. At the
lowering of the temperature $t_c(T)$ decreases, and at
\begin{equation}\label{2}
t_c(T_c) \ll T_c/E_F
\end{equation}
SCT occurs according to the scenario Kats at $T_c >T_{2D}$ [16]: from
the beginning two-dimensional crossovers 0D SCF$ \rightarrow$ 2D SCF
$\rightarrow$ 3D SCF, and then occurs three-dimensional SCT with coherent
charge transfer along $c$-axis [13-15,17]. As we can see from (1) and (2)
[10]
\begin{equation}\label{3}
T_{2D} < T_c \le \frac{\overline{\xi}_c ^{2}  E_F T_{2D}}
{\overline{\xi}_c ^{2}E_F - \overline{\xi}_{ab} ^{2}T_{2D}}
\end{equation}

Thus, SCT has two-dimensional character with limited region of
three-dimensional SCF. For example, as it was shown from the analysis
of the resistivity measurements in single crystal Bi-2212 with
$T_c \sim 80 K$ [14], the region of (0D SCF + 2D SCF) $\sim 120 K$, and
the region 3D SCF $\sim 10 K$.

4. For HT superconductivity the conclusion about the decisive role of
JT SP qualitatively comes to an agreement with the dependence on the
doping of the part of each carriers type relatively of total carriers
number: in Ref.[4] it was shown that the value of
$T_c $ ( at the "hole-JT SP" pairing   $T_c \sim n_h \times n_p$ )
amount to $T_{c, max}$ at the concentration 0.15 on ion $Cu_{+2}$ when  the
part of polarons is $\sim0.6$, and the part of holes is $\sim 0.4$.

The measurements of the resistivity also evidence about the coexistence
of carriers and local pairs "hole - JT SP" [13-15,17] in PGS. The
studying of fluctuational conductivity [15,17] shown that interactions of
fluctuational pairs with carriers are weak, and the contributions
of 0D, 2D and 3D SCF into conductivity are been identified.

Once more convincing example of the coexistence of the holes, polarons
and local pairs "hole - JT SP" in PGS is the observation of a doublet
structure of two-magnon absorption band at 2.145 eV and 2.28 eV
in a metal films $YBa_2 Cu_3 O_{6+x} (x\sim 0.85)$, $T_c = 87.4K)$ [18].
The doublet structure was observed at $T<T^{\ast}$ for PGS and for
superconducting state (SCS). Its first component with energy
$\omega\approx \Delta _{CT}+3J$ is identical to that
which was observed in doped AF with $x=0.3$, and was caused by polaron
two-magnon absorption at interband transition of polaron (here
$J\sim0.13 eV$  is the exchange energy, $\Delta _{CT}$ is transfer energy).
 As it is known, in metal phase of the sample at $T>T^{\ast}$ this
component is not observed, and in Ref.[18] at lowering the temperature
it is observed only at $T<T^{\ast}$ in both PGS and SCS. Second component
of the doublet with $\omega\approx \Delta _{CT}+4J$ is observed only at
 $T<T^{\ast}$. We conjecture that this component was caused by the
fulfilment of the condition of the "triple
resonance" at polaron two-magnon absorption, similar that at Raman
scattering for undoped AF [19-20]. JT SP absorbs of photon (with
energy $\omega$) and transfer into valence band. Two transfers of the
charge with energy $t$ (between $Cu^{+2}$ and oxygen ion within JT SP
complex, there and back) lead to the radiation of two magnons with
the frequencies $\Omega_q, \Omega_{-q}$ at the  resonance condition
\begin{equation}\label{4}
\omega=\Delta_{CT}+2t+\Omega_q  + \Omega_{-q}.
\end{equation}
For UD HTS at $T>T^{\ast}$ this condition is not practicable with taken
into account the interactions between holes and polarons, but both
incoherent holes and polaron two-magnon absorption lead to an essential
asymmetry and big width of right hand of two-magnon absorption band (the
so-called background): for $YBa_2 Cu_3 O_{6.1}$ up to $\omega\sim  3eV$
[18-20]. At $T<T^{\ast}$ holes and part of JT SP $n_p ^{\ast} \sim n_h $
are in pairing state, and for unpaired part of polarons $\sim (n_p
-n_h)|_{T<T^{\ast}}$ resonance condition (4) takes place in both PGS and SCS.

The observation of doublet structure of two-magnon absorption band
evidences about the existence of the polarons in PGS and SCS, and
about  essential charge heterogeneity of SCS ( the same as for PGS).
A doublet structure of two-magnon absorption band for SCS is an indirect
evidence of $d$-wave symmetry of superconducting order parameter for UD HTS
as well: at $T<T_c$ some part of unpaired polaron $\sim(n_p-n_h)|_{T<T_c}$
percolates through the direction of wave vector where the order parameter
is equal zero.

To summarize, in this paper we show that for high temperature
 superconductivity of cuprates the coexistence in normal state
of holes and Jahn-Teller small polarons is  fundamentally
important, but the decisive role belong the latter
which at strong Jahn-Teller interactions (for
Cu$^{+2}$ this energy is $\sim 1.2 eV$) lead to polaron shift of
 energy, and to the possibility of the compensation of Coulomb repulsion
between polaron and hole, at which  on-site attraction of holes and
polarons in $Cu O_2$ planes and BCS pairing takes place.

References

1. J.Orenstein, G.A.Thomas, A.J.Millis et al. Phys.Rev. B42, 6342, 1990.

2. S.Uchida, T.Ido,H.Takagi et al., Phys.Rev. B40, 7942, 1991.

3. X.-X. Bi and P.C.Eklund, Phys.Rev. Lett. 70, 2625, 1993.

4. K.A.Muller, G.-M.Zhao, K.Konder, and H.Keller, J.Phys.: Condens. Matter,
10, L291 (1998).

5. K.H.Hock, H.Nikisch, and H.Thomas, Helv.Phys.Acta 56, 237
(1983).

6. D.Mihailovic, J.Demsar, B.Podobnik, V.V.Kabanov, J.E.Ivetts,
G.A.Vagner, and L.Mechin, J.Superconductivity, 12, 33 (1999).

7. A.F.Barabanov, L.A.Maksimov, A.V.Miheenkov, JEThP Lett. 74, 674 (2001). 

8. E.I.Kudinov,  44, 667 (2002).

9. B.I.Kochelaev, J.Superconductivity, 12, 53 (1999).

10. G.G.Sergeeva, Voprosy Atomnoj Nauki i Tehniki, 2001, N6, p.375

11. D.N.Basov, C.C.Homes, E.J.Singley et al.  Phys.Rev. B 63, 134514, (2001).

12. I.O.Kulik, A.G.Pedan, Fizika Nizkih Temperatur (Russin) 14, 700 (1988).

13. G.Balestrino, M.Marinelli, and E.Milani, A.A.Varlamov, and L.Yu,
Phys.Rev. B 47,  8936 (1993).

14. G.G.Sergeeva, V.Yu.Gonchar, A.V.Voytsenya, Fizika Nizkih Temperatur
(Russin)    27, 634, (2001).

15. A.L.Solovjov, H-U.Habermeier, and T.Haage, Low Temp.Physics (FNT) 28,
N2, 144 (2002)

16. E.I.Kats, JETPH (Russin) 56, 1675 ( 1965).

17. A.L.Solovjov, H-U.Habermeier, and T.Haage, Low Temp.Physics (FNT) 28,
N1, 24 (2002). 

18. V.V.Eremenko, V.N.Samovarov, V.L.Vakula, M.Yu.Libin, S.A. Uyutnov,
V.M.Rashkovan, Low Temp.Physics (FNT) 27, N 11, 1327 (2001)

19. A.Chubukov, D.Frenkel, Phys.Rev. Lett. 74, 3057, 1995.

20. D.K.Morr, A.V.Chubukov, A.P.Kampf, G.Blumberg, Phys.Rev. B 54, 11930,
(1996).
                                                                                                                                                                                                                                                                                                                                                                                                                                  
\end{document}